\documentclass[12pt,preprint]{aastex63}
\usepackage{natbib}
\shorttitle{Wind Lines of Rapid Rotators} 
\shortauthors{Shepard et al.} 
 
\begin{document} 
 
\received{September 27, 2019} 
\accepted{November 20, 2019} 

\title{HST/COS Spectra of the Wind Lines of VFTS 102 and 285} 
 
\correspondingauthor{Katherine Shepard}

\email{shepard@astro.gsu.edu, gies@chara.gsu.edu, lester@astro.gsu.edu, lwang@chara.gsu.edu}
\email{zxg124@psu.edu,  L.Kaper@uva.nl, A.deKoter@uva.nl, hugues.sana@kuleuven.be}

\author[0000-0003-2075-5227]{Katherine Shepard}
\affiliation{Center for High Angular Resolution Astronomy and  
 Department of Physics and Astronomy,\\ 
 Georgia State University, P. O. Box 5060, Atlanta, GA 30302-5060, USA}

\author[0000-0001-8537-3583]{Douglas R. Gies}
\affiliation{Center for High Angular Resolution Astronomy and  
 Department of Physics and Astronomy,\\ 
 Georgia State University, P. O. Box 5060, Atlanta, GA 30302-5060, USA}

\author[0000-0002-9903-9911]{Kathryn V. Lester}
\affiliation{Center for High Angular Resolution Astronomy and  
 Department of Physics and Astronomy,\\ 
 Georgia State University, P. O. Box 5060, Atlanta, GA 30302-5060, USA}

\author[0000-0003-4511-6800]{Luqian Wang}
\affil{Center for High Angular Resolution Astronomy and  
 Department of Physics and Astronomy,\\ 
 Georgia State University, P. O. Box 5060, Atlanta, GA 30302-5060, USA}

\author[0000-0002-0951-2171]{Zhao Guo}
\affiliation{Center for Exoplanets and Habitable Worlds, Department of Astronomy and Astrophysics, \\
Pennsylvania State University, 525 Davey Lab, University Park, PA 16802, USA}

\author{Lex Kaper}
\affiliation{Anton Pannekoek Institute for Astronomy, University of Amsterdam, \\ 
Science Park 904, 1098 XH Amsterdam, The Netherlands}

\author{Alex De Koter}
\affiliation{Institute of Astrophysics, KULeuven, Celestijnenlaan 200 D, 3001 Leuven, Belgium; \\ 
Anton Pannekoek Institute for Astronomy, University of Amsterdam, \\ 
Science Park 904, 1098 XH Amsterdam, The Netherlands} 

\author[0000-0001-6656-4130]{Hugues Sana}
\affiliation{Institute of Astrophysics, KULeuven, Celestijnenlaan 200 D, 3001 Leuven, Belgium}


 
\begin{abstract} 
Rapid rotation in massive stars imposes a latitudinal variation in the 
mass loss from radiatively driven winds that can lead to enhanced 
mass loss at the poles (with little angular momentum loss) and/or equator
(with maximal angular momentum loss).  Here we present an examination 
of the stellar wind lines of the two O-type stars with the fastest known 
equatorial velocities, VFTS~102 ($V\sin i = 610 \pm 30$ km~s$^{-1}$; O9:~Vnnne+) 
and VFTS~285 ($V\sin i = 609 \pm 29$ km~s$^{-1}$; O7.5~Vnnn)
in the Large Magellanic Cloud.  Ultraviolet spectra of both stars were obtained with 
the {\it Hubble Space Telescope} Cosmic Origins Spectrograph.  The spectrum of VFTS~285 
displays a fast outflow in \ion{N}{5} and a much slower wind in \ion{Si}{4},
and we argue that there is a two-wind regime in which mass loss is 
strong at the poles (fast and tenuous wind) but dominant at the equator (slow and dense winds).
These ions and wind lines are not present in the spectrum of the cooler star VFTS~102,
but the double-peaked H$\alpha$ emission in its spectrum implies 
equatorial mass loss into a circumstellar disk.  The results suggest
that in the fastest rotating O-stars, most mass is lost as an equatorial outflow, 
promoting angular momentum loss that contributes to a spin down over time. 
\end{abstract} 
 

 
\NewPageAfterKeywords

 
\setcounter{footnote}{0} 
 
\section{Introduction}                              
 
Massive stars are hot, luminous objects that experience mass loss 
by radiatively-driven stellar winds (e.g., \citealt{Puls2008}).  
Many massive stars are also rapid rotators (e.g., \citealt{Ramirez2013}), 
however despite this commonality, the manner in which rotation affects 
the wind geometry and mass loss properties of the star remains a mystery. 

\citet{Friend1986} and \citet{Pauldrach1986} began to study this problem utilizing
calculations that included the effective reduction in gravitational acceleration.
This characteristic is a result of the rotational centrifugal acceleration which reaches
maximum effect at the equator.
These models and a subsequent study by \citet{Bjorkman1993}
of the crossing wind trajectory lines over the equator led to the 
prediction that the effective mass loss rate would increase from pole to 
equator.  However, very rapid rotators are expected to have cooler 
photospheres closer to their equatorial regions (due to gravity 
darkening), and further investigations that accounted for cooler 
equatorial zones and nonradial line forces found that, in fact, the local mass loss 
rate attains a minimum at the equator \citep{Owocki1996, Maeder2000, Pelupessy2000} and 
that the strongest winds occur at the poles driving bipolar outflows 
\citep{Dwarkadas2002,vanBoekel2003,Puls2008}.  Detailed wind models by \citet{Muller2014}
confirm that the polar outflows dominate the mass loss, and 
because the polar gas has little angular momentum, massive stars 
would lose matter but retain angular momentum, leading to a slower reduction 
in rotation over time.  

Another important factor is the temperature of the photospheric gas and 
the varieties of ionic transitions that act as wind-driving opacity sources. 
\citet{Vink1999} showed that as temperatures decline below approximately 25000~K
the ionization balance of Fe atoms shifts from \ion{Fe}{4} to \ion{Fe}{3},
and the multitude of line transitions of the latter act in the winds 
of B-supergiants to increase dramatically the mass loss rate on the 
cooler side of a {\it bi-stability jump}, a point of discontinuity in 
the properties of stellar winds.  
Such cooler temperatures could exist in the 
equatorial regions of rapidly rotating stars due to gravity darkening, 
and \citet{Lamers1991} showed that in such cases mass loss might be enhanced 
over these regions (confirmed for differing circumstances by \citealt{Owocki1998}, 
\citealt{Maeder2000}, \citealt{Pelupessy2000}, and \citealt{Madura2007}).   
This trend was recently and vividly illustrated in wind calculations by 
\citet{Gagnier2019a}, who presented models for the latitudinal variation
in wind strength for realistic, two-dimensional models of rapidly 
rotating O-stars \citep{EspinosaLara2011}.  They found that for very 
rapid rotation the local temperature may drop below the bi-stability 
jump causing a large increase in the local mass loss rate at the equator.  The star then 
enters a two-wind regime with a fast, low density wind at high latitudes
and a slow, very dense outflow near the equator.  In this case, the 
latter component becomes the dominant source of mass and angular momentum loss. 

Given the two very different outcomes of the models (polar versus equatorial 
mass loss), we must seek guidance from observational results for luminous, rapidly rotating stars.
The far ultraviolet spectra of the O-stars are marked by three major resonance doublets, 
\ion{N}{5} $\lambda\lambda 1238, 1242$, \ion{Si}{4} $\lambda\lambda 1393, 1402$, 
and \ion{C}{4} $\lambda\lambda 1548, 1550$, which are generally formed 
by scattering in the stellar wind \citep{Howarth1989}.  These lines appear
as P~Cygni profiles with blue-shifted absorption (from the column of gas carried by the wind 
projected against the star) and red-shifted emission (from the halo of scattered light 
around the star), and the line strengths and outflow velocities 
generally vary according to temperature and gravity associated with the spectral 
classification \citep{Walborn1985, Howarth1989}.  However, the P~Cygni lines 
of several rapid rotators show unusual properties for their classifications. 
The first case is that of the rapid rotator HD~93521 (O9.5~V; $V\sin i=341$ km~s$^{-1}$).
\citet{Howarth1993} found evidence of both a strong, low velocity component in the lines
of \ion{Si}{4} and \ion{C}{4} and a weaker but high velocity component in 
\ion{N}{5} and \ion{C}{4}.  They interpreted this as the result of a latitudinal 
dependence in the wind ranging from fast at the poles to slow at the equator. 
\citet{Bjorkman1994} analyzed the same features and argued for an equatorial wind enhancement 
through a detailed comparison of the wind line shapes with models (see below). 
\citet{Massa1995} came to a similar conclusion based upon 
the strong absorption and weak emission of the lower ionization species that probably 
form in a denser equatorial region that is seen directly in projection against 
the stellar disk. 
\citet{Massa1995} also considered the second example of a rapid rotator, $\zeta$~Oph 
(HD~149757; O9.5~Vnn; $V\sin i = 348$ km~s$^{-1}$).  The \ion{N}{5} and \ion{C}{4}
lines in its spectrum have strong emission but relatively weak absorption components, 
which implies that the regions containing these ions do not fully cover the 
stellar disk and are absent in a dense equatorial zone with lower ionization states 
seen in partial projection against the star.  
\citet{Prinja1997} described a third case, $\gamma$~Arae
(HD 157246; B1~Ib), that has a relatively large projected rotational velocity 
for a supergiant, $V\sin i = 230$ km~s$^{-1}$.  The \ion{Si}{4} lines are strongly
in absorption with little emission in the same way as they appear in the spectrum
of HD~93521.  Furthermore, there are two sets of migrating Discrete Absorption Components
that reach differing outflow speeds, suggestive of an association with both the slower equatorial
zone and faster polar region.  The last example is the star BI~208 in the Large 
Magellanic Cloud (O6~V((f)); $V\sin i = 240$ km~s$^{-1}$; \citealt{Massey2009}). 
\citet{Massa2003} noted that the \ion{C}{4} absorption is relatively weak in 
this star's spectrum and resembles that of HD~93521, suggesting a wind asymmetry.

The modifications of the appearance of the stellar wind lines in rapid rotators 
were explored in a seminal paper by \citet{Bjorkman1994}.  They developed 
analytical models for line formation in an expanding wind that varies with 
stellar co-latitude.  Their model posits a wind with a terminal velocity 
and local mass loss rate that experiences a transition from a fast polar flow 
to a slower equatorial flow at an angle corresponding to the equatorial disk
half-width.  The line profiles are
found by integration of scattered flux in the wind using an escape probability 
method and an analytical representation for the wind ionization as a function of 
distance from the star.  The resulting wind line profiles depend upon parameters
that describe the geometry and outflows (stellar equatorial velocity, polar 
and equatorial wind terminal velocities and mass loss rates, the equatorial zone 
half-width opening angle, and the inclination between the spin axis and observer) 
and the ionization state (different for each line).  \citet{Bjorkman1994}
present a montage of model profiles for variations in each parameter to 
demonstrate how these individually influence the final shape of the P~Cygni
wind profile.  Their main result is that the two-component wind model creates 
an absorption profile shape with a flat ``shelf'' connecting the deep, low 
velocity component to the shallower, high velocity component that extends to 
the polar terminal velocity.  Their fits of the models to the observed profiles 
in the spectrum of HD~93521 indicate that the equatorial density enhancement 
has a small opening angle, a local mass rate about ten times larger than the 
polar value, and is viewed almost directly edge-on (inclination of $90^\circ$). 
These models offer important guidance about the interpretation of the P~Cygni 
line profiles in other rapid rotators.

We now have the opportunity to explore the winds of extremely rapid rotators 
in two cases discovered in the vicinity of the Tarantula Nebula complex in the 
Large Magellanic Cloud.  The VLT-FLAMES Tarantula Survey (VFTS; \citealt{Evans2011}) 
was a large-scale survey to obtain optical spectroscopy of over 800 
massive stars in the 30~Doradus region.  Among the many remarkable discoveries 
made in the survey was the identification of two very rapid rotators.  
\citet{Dufton2011} described the spectrum of \object{VFTS 102}, a late-type 
O-star with $V\sin i = 610 \pm 30$ km~s$^{-1}$ \citep{Ramirez2013}, the largest projected 
rotational velocity ever measured for this class of stars.  
A second fast rotator was discovered by \citet{Ramirez2013}, 
\object{VFTS 285}, with $V\sin i = 609 \pm 29$ km~s$^{-1}$.  These two fast rotators are 
conspicuous outliers in the $V\sin i$ distribution of the VFTS sample, where the majority
of single O-type stars have a $V\sin i$ between $40$ and $120$ km~s$^{-1}$
(see Fig.~5 and Fig.~11 in \citealt{Ramirez2013}).  The cause of such rapid rotation
in these two stars is unknown.  \citet{Dufton2011} discuss the possible origin of VFTS~102 
as a star spun-up by mass accretion in a binary system that subsequently broke apart upon 
the supernova explosion of the companion that created the nearby X-ray pulsar, 
PSR J0537-6910.  Alternatively, \citet{Jiang2013} and \citet{deMink2014} describe how a 
short-period binary might evolve into a contact system that eventually merges to become a 
single, rapidly rotating star.  

These two extremely rapid rotators are ideal targets to explore the wind properties 
in the case of close to critical rotation, where the equatorial velocity is 
nearly equal to the Keplerian orbital velocity.  Here we present spectra of the \ion{N}{5} and 
\ion{Si}{4} profiles of VFTS~102 and 285 obtained with the {\it Hubble Space Telescope}
(HST) and Cosmic Origins Spectrograph (COS).  The observations and data reduction 
are discussed in Section 2.  The wind lines contain components formed in the 
interstellar medium (ISM) of the Galaxy and LMC, and in Section 3 we describe our 
method to isolate and remove these ISM components.  Then in Section 4 we investigate 
the wind line morphology in the context of the wind lines of Galactic rapid rotators
that were observed with the {\it International Ultraviolet Explorer} (IUE). 
Our results are summarized in Section 5. 

 
\section{HST/COS Spectroscopic Observations}        

The Cosmic Origins Spectrograph (COS) is a high dispersion spectrograph 
that was designed to record the UV spectra of faint point sources
\citep{Green2012, Fischer2019}.  The observations reported here were 
made during Cycle~23 under program GO-14246.  The HST/COS observations 
of VFTS~102 were obtained over three orbits on 2017 January 1, and 
those of the brighter VFTS~285 were made in a single orbit on 2016 April 10.  
These far-UV spectra were all made with the G130M grating in order to record the spectrum 
over the range from 1150 to 1450 \AA\ with a spectral resolving power of 
$R=\lambda/\triangle\lambda=18000$.  There are two COS detectors
that are separated by a small gap, therefore the VFTS~102 spectra were made 
at three slightly different central wavelengths (1300, 1309, and 1318~\AA ) 
in order to fill in the missing flux.  In each of these settings, 
four sub-exposures were obtained at four {\tt FP-POS}, or focal plane offset positions, in 
order to avoid fixed-pattern problems.   The VFTS~285 spectra were made in 
the same way except only two central wavelength positions were selected, 
1300 and 1318~\AA , due to orbital time restrictions. 

The observations were processed with the standard COS pipeline 
to create wavelength and flux calibrated spectra as {\tt x1d.fits} files 
for each central wavelength arrangement \citep{Rafelski2013}.  
The sub-exposures were subsequently merged onto a single 
barycentric wavelength grid using the IDL procedure 
{\tt coadd\_x1d.pro}\footnote{http://casa.colorado.edu/$^\sim$danforth/science/cos/coadd\_x1d.pro}
\citep{Danforth2010}.  Finally, both spectra were 
transformed to a uniform grid with a $\log \lambda$ pixel 
spacing equivalent to a Doppler shift step size of 2.60 km~s$^{-1}$
over the range from 1150 to 1440 \AA .  The coadded spectra have
a signal-to-noise ratio of S/N = 5 per pixel in the central, best exposed regions.

The spectra are composed of very broad and blended photospheric lines, very strong Ly$\alpha$
absorption, the \ion{N}{5} and \ion{Si}{4} wind lines, and numerous sharp interstellar lines.  
For the purposes of this work, the spectra were flux normalized by dividing by the mean 
flux in the region from 1340 to 1440 \AA ~($3.45\times 10^{-15}$ and $2.11\times 10^{-14}$
erg~cm$^{-2}$~s$^{-1}$~\AA $^{-1}$ for VFTS~102 and VFTS~285, respectively). 
The shape of the observed continuum flux distribution is unaltered. 
We will defer a discussion of the rotational line broadening of the photospheric lines 
for a subsequent paper and concentrate on the wind lines in this work. 

 
\section{Line Components from the Interstellar Medium} 

Inspection of the HST/COS spectra of VFTS~102 and 285 revealed that the wind lines
(especially \ion{Si}{4}) are inscribed with narrow interstellar components formed in both the Galaxy and LMC. 
In order to analyze the wind lines originating from the star itself, we need to remove
the ISM components. We approached this problem by searching the HST archive 
for other COS and G130M grating spectra of luminous stars in the LMC in order to identify
the general shapes and strengths of the ISM absorption components.  Table~1 lists 
nine other stars that have spectral observations comparable to those of VFTS~102 and 285.
The columns of Table~1 list the target name used in the archive, other name, spectral 
classification and source, $B-V$ color index 
\citep{Massey2002, Zaritsky2004, Bonanos2009, Evans2011, Zacharias2013}, 
the number of available COS spectra, the HST program 
number and principal investigator, and whether or not (Y=yes or N=no) the target spectra 
were used in our scheme to remove the ISM lines.  
Similar data are given for the two primary targets at the bottom of Table~1.  
These spectra were reduced to 
the same wavelength grid using the methods applied to the VFTS~102 and 285
spectra, and where multiple spectra were available, they were co-added to form a single, high 
S/N spectrum.  We further smoothed the spectra by convolution with an 11 pixel 
boxcar function, reducing the spectral resolution to $R=\lambda / \Delta\lambda = 10000$.
Finally we normalized the spectra to the median average of the flux in the 1397 to 1402 \AA\ range,
leaving the general shape of the spectral energy distribution the same. 
For our purposes, we restricted this set to four target spectra that 
display very broad stellar features in the vicinity of \ion{Si}{4} $\lambda\lambda 1393, 1402$, 
so that the narrow ISM lines are readily identified by their sharp appearance.  
These four spectra are shown in Figure~1, and in all these cases the ISM components
appear as narrow absorptions superimposed upon broad P~Cygni type profiles. 
Note that the four selected stars and two primary targets all have similar observed and 
intrinsic color indices (with the possible exception of VFTS~102 that may 
appear redder because of the flux contribution from the red continuum of 
its circumstellar disk; see Section 4.3 below), so we expect that all have similar 
interstellar column densities. 

\placetable{tab1}      
\begin{deluxetable*}{lllcccclc}
\tabletypesize{\scriptsize}
\tablenum{1}
\tablecaption{Other LMC Stars Observed With COS G130M\label{tab1}}
\tablewidth{0pt}
\tablehead{
\colhead{Name} & 
\colhead{Other Name} & 
\colhead{Sp. Class.} & 
\colhead{Ref.} &
\colhead{$B-V$} &
\colhead{No. Spec.} & 
\colhead{HST Prog.} &
\colhead{PI} & 
\colhead{Used}
}
\startdata
AA$\Omega$ 30Dor 187 & UCAC3 42-30814        & O6 (f)np        & 1  & $-0.18$ & \phn 4 & 14683 & J.-C. Bouret   & Y \\
AA$\Omega$ 30Dor 333 & CPD--69$^\circ$ 471   & O2-3 (n)fp      & 1  & $+0.10$ & \phn 8 & 14683 & J.-C. Bouret   & Y \\
AA$\Omega$ 30Dor 368 & UCAC3 42-33014        & O7.5 (f)np      & 1  & $-0.12$ & \phn 4 & 14683 & J.-C. Bouret   & N \\
HD 38029             & Brey 67               & WC4 + OB        & 2  & $+0.15$ & \phn 1 & 12581 & J. Roman-Duval & N \\
SK-68$^{\circ}$ 129  & W61 27-56             & B1 I            & 3  & $+0.03$ & \phn 2 & 12581 & J. Roman-Duval & Y \\
SK-68$^{\circ}$ 140  & VFTS 696              & B0.7 Ib-Iab Nwk & 4  & $+0.10$ & \phn 1 & 12581 & J. Roman-Duval & N \\
SK-68$^{\circ}$ 155  & UCAC2 1934920         & O8 Ia           & 5  & $-0.01$ & \phn 1 & 12581 & J. Roman-Duval & Y \\
SK-69$^{\circ}$ 279  & UCAC2 1677771         & O9.2 Iaf        & 6  & $-0.06$ & \phn 1 & 12581 & J. Roman-Duval & N \\
VFTS 352             & OGLE BRIGHT-LMC-ECL-9 & O4.5 V + O5.5 V & 7  & $-0.10$ &     16 & 13806 & H. Sana        & N \\
\tableline
VFTS 102             & UCAC2 1803231         & O9: Vnnne+      & 8  & $+0.35$ & \phn 1 & 14246 & D. Gies        & \nodata \\
VFTS 285             & \nodata               & O7.5 Vnnn       & 8  & $-0.06$ & \phn 1 & 14246 & D. Gies        & \nodata \\
\enddata
\tablecomments{References for classifications:  1.\ \citet{Walborn2010}; 2.\ \citet{Neugent2018};
 3.\ \citet{Massey2000}; 4.\ \citet{Evans2015}; 5.\ \citet{Misselt1999}; 6.\ \citet{Gvaramadze2018};
 7.\ \citet{Abdul-Masih2019}; 8.\ \citet{Walborn2014}.}
\end{deluxetable*}

\begin{figure*}[ht!]
\begin{center} 
{\includegraphics[height=8cm,angle=0]{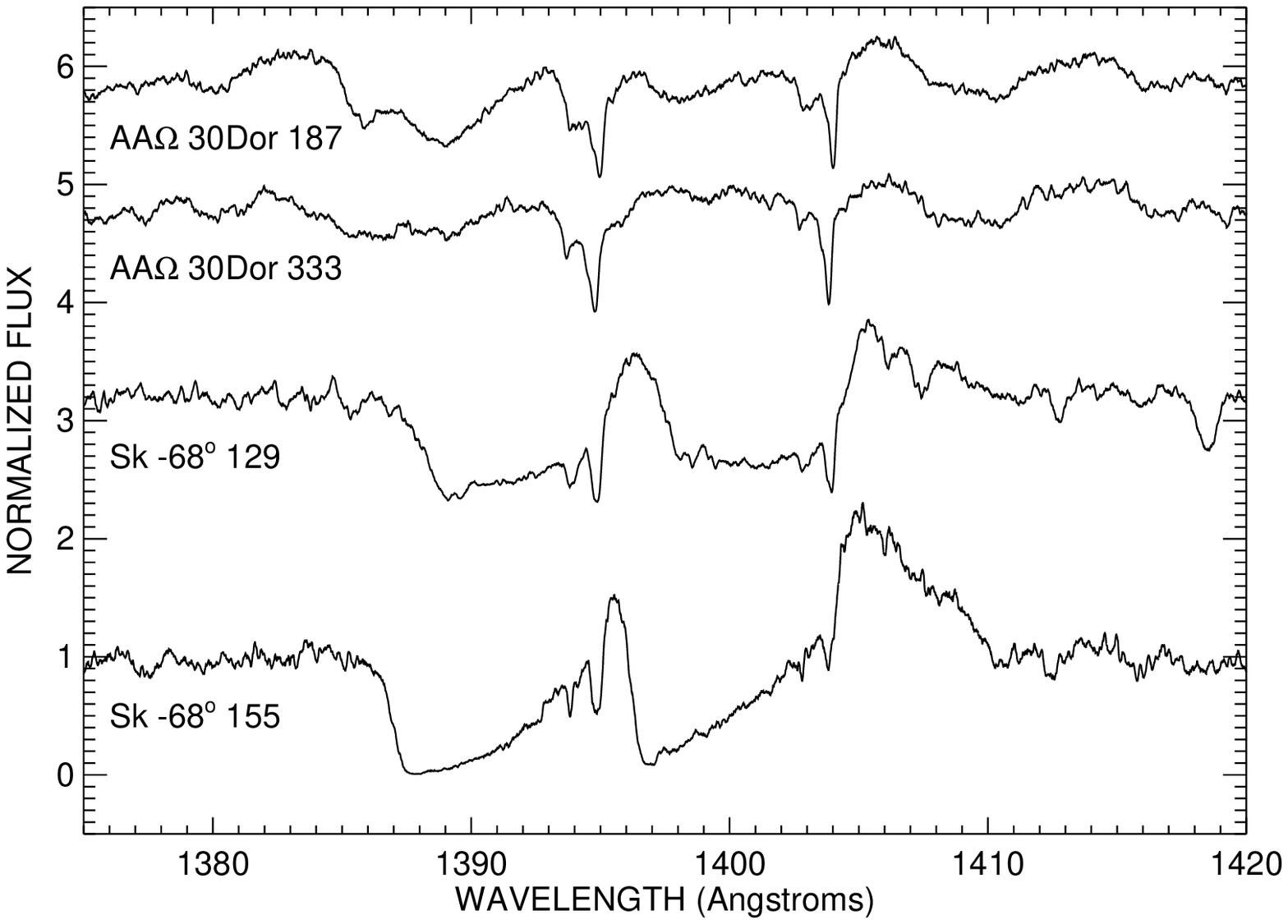}} 
\end{center} 
\caption{HST/COS spectra of LMC stars in the vicinity of \ion{Si}{4} $\lambda\lambda 1393, 1402$. 
The ISM components from the Galaxy (left) and LMC (right) are seen as
narrow absorptions in both components of the doublet.  The normalized fluxes are offset 
vertically for clarity by $+5.3$, $+4.2$, $+2.2$, and $0$ from top to bottom, respectively. 
} 
\label{fig1}
\vspace*{\floatsep}
\placefigure{fig2}     
\begin{center} 
{\includegraphics[height=15cm,angle=90,trim= 1.in 1.5in 1.0in 0.0in]{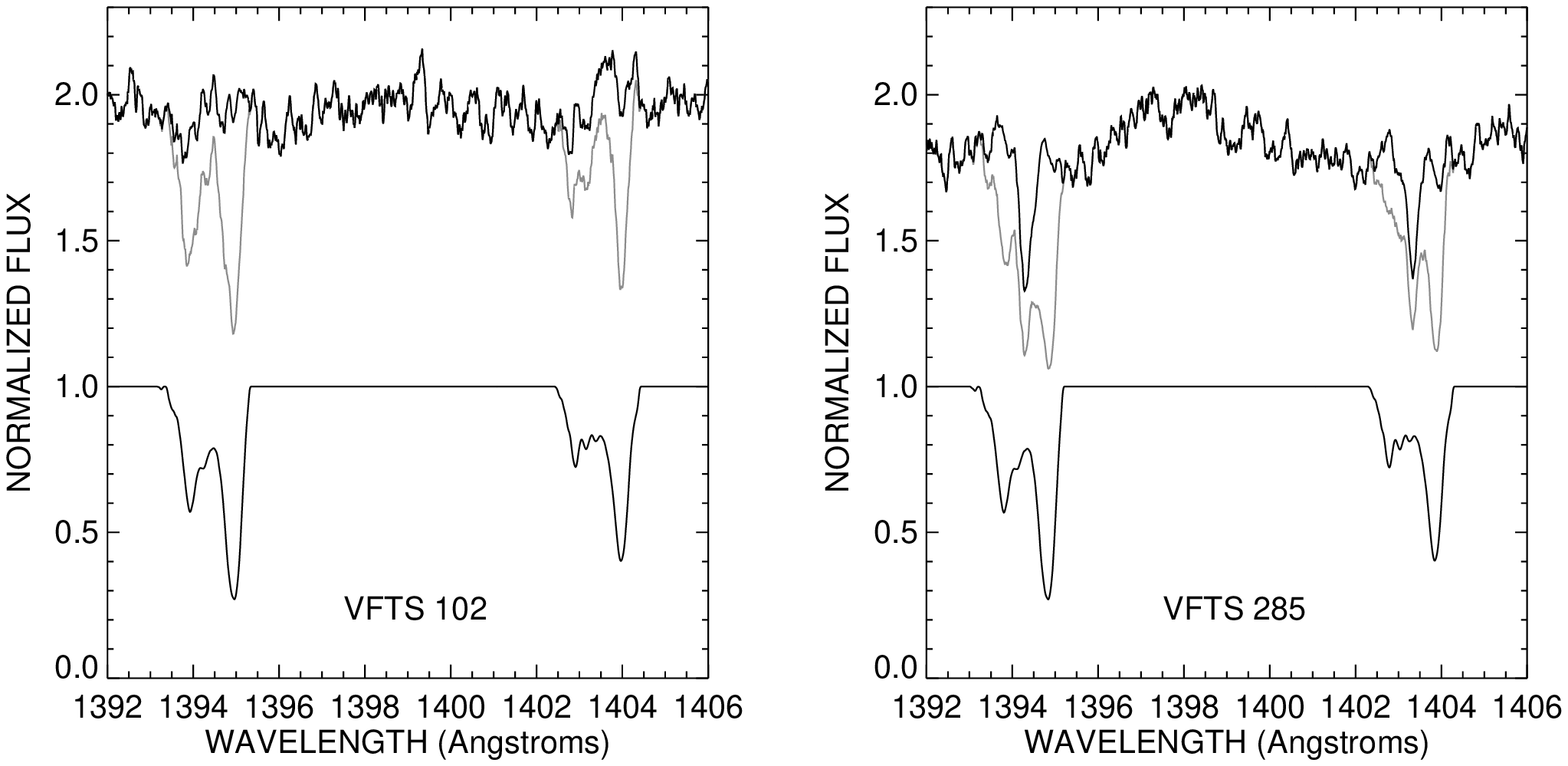}} 
\end{center} 
\caption{HST/COS spectra of VFTS~102 and VFTS~285 in the vicinity of \ion{Si}{4} 
$\lambda\lambda 1393, 1402$. The lower solid line shows the expected ISM components for the doublet.
The upper lines (offset by $+1$) depict the observed spectra with both the stellar and interstellar 
components (gray) and with the ISM components removed (black). 
} 
\label{fig2}
\end{figure*}

\placefigure{fig1}     


We used these four spectra to form an average ISM component spectrum\footnote{
Note that in principle this averaging should be done by optical depth 
scaling by first calculating the logarithms of the residual intensities, averaging these, 
and then forming the exponent of the result. However, because all the ISM components 
have similar fractional depths for these four stars, the simple average suffices for our purposes.}
for both VFTS~102 and 285. Each spectrum was shifted by a few pixels allowing the deepest LMC absorption 
component to align with the same feature in the VFTS target spectra. Then all four were averaged 
together to form a single spectrum.  Next we interpolated across the span of the 
interstellar components by making parabolic fits of the immediately adjacent line regions. 
The ISM components were then extracted by dividing by the parabolic fits.  
The final isolated interstellar lines are shown in the lower portion of Figure~2.  
The \ion{Si}{4} $\lambda\lambda 1393, 1402$ profiles of VFTS~102 and 285 are plotted 
in Figure~2 after smoothing to the same resolving power of $R= 10000$.  The gray line 
corresponds to the observed spectrum and the black line shows the net profiles after 
subtraction of the ISM lines.  In principle, the interstellar absorptions 
remove a fraction of the flux, therefore technically we should divide rather than subtract their profiles, 
however we found that subtraction better preserved the S/N properties of the observed spectra 
and appeared to succeed in removing most of the interstellar components.  We note that the 
small shifts we applied to align the LMC components may not be appropriate for the 
Galactic components along the line of sight, but because the Galactic components are 
relatively weak, this inconsistency has only minor consequences for the subtracted spectra. 

The results of ISM removal for the \ion{Si}{4} $\lambda\lambda 1393, 1402$ lines (Fig.~2) 
show that the spectrum of VFTS~102 has little to no wind contribution in this transition. 
On the other hand, the ISM-subtracted spectrum of VFTS~285 shows a strong net absorption for 
both members of the doublet that is Doppler shifted to a position between the ISM components 
of the Galaxy and LMC.  Such strong features at these wavelengths are not seen in the 
spectra of any of the other luminous LMC stars listed in Table~1. Therefore, we conclude that 
this feature has its origin in the star itself.  We consider a wind origin for this 
feature in the next section. 

 
\newpage

\section{Wind Profiles of Rapid Rotators}           
 
\subsection{Comparison to a Sample from IUE Spectroscopy} 

Our intent is to place the appearance (or absence) of the wind profiles in the 
spectra of VFTS~102 and 285 into the context of the observational properties of 
wind lines among O-stars in general and rapid rotators in particular. 
The strengths of the absorption troughs and emission peaks of the wind lines 
vary as a function of mass loss rate (dependent primarily on luminosity and metallicity), 
velocity law (often set by exponent $\beta$ to determine the velocity as a function 
of radial distance $r$ by $v(r)=v_\infty (1 - R_\star/r)^\beta$), 
terminal velocity (related to escape velocity), and temperature (related to 
characteristic ionization fractions).  Consequently, the appearance
of the wind lines is closely related to the spectral classification (associated 
with photospheric temperature and gravity), and there are clear trends with both 
spectral type and luminosity class.  The trends in the primary P~Cygni wind 
lines of \ion{N}{5} $\lambda\lambda 1238, 1242$, \ion{Si}{4} $\lambda\lambda 1393, 1402$, 
and \ion{C}{4} $\lambda\lambda 1548, 1550$ for Galactic O-stars are displayed in Figure~2 
of the paper by \citet{Howarth1989} and additionally in the atlas by \citet{Walborn1985}.  
The \ion{N}{5} feature grows in strength and extent (towards larger terminal velocity) 
from near invisibility among late-type O-stars to dominance among early-type
main sequence stars. Similar trends are observed in giant O-stars, while in  
supergiants, the \ion{N}{5} lines are strong among all spectral subtypes. 
The \ion{Si}{4} line, on the other hand, generally shows only photospheric 
components (or blends with ISM components) among the O-dwarfs, and wind profiles
are only found among the more luminous O-stars (particularly the cooler ones). 
\citet{Crowther2016} present a similar spectral montage for O-stars in the LMC
(see their Figures A1 to A11). Although many of the same general trends are 
present in the LMC O-stars, the wind lines tend to be somewhat weaker than found in 
their Galactic counterparts due to the lower metallicity of the LMC and reduced 
wind-driving line opacities.  

The wind lines of rapidly rotating Galactic O-stars show a great diversity of 
morphology that often departs from the general trends \citep{Howarth1993, Massa1995}.
Nevertheless, there are patterns among the spectral lines of these rapid rotators 
that can help place those of the LMC stars in the context of spectral classification 
and rotational velocity.  For this purpose, we collected the available high dispersion 
spectra of extremely rapid rotators from the IUE archive for a direct comparison
of their wind lines with those of VFTS~102 and 285.  The stars selected have spectral 
subtypes from O7 to O9, and are all likely very rapid rotators.  Most have 
large measured projected rotational velocity $V\sin i$ that is also indicated by one or more ``n'' 
labels attached as a suffix to the classification to indicate very broad absorption lines.
Several others have an ``e'' suffix denoting Balmer emission that is usually 
associated with a circumstellar disk, as found among the rapidly 
rotating Be stars \citep{Negueruela2004}.  Selected stellar parameters for the sample of 
seven Galactic rapid rotators plus those of VFTS~102 and 285 are summarized in Table~2.
The columns list name, number of Short Wavelength Prime, high resolution spectra from the IUE archive, 
spectral classification and reference, radial velocity and reference, projected 
rotational velocity and reference, and wind terminal velocity and mass loss
rate from the work of \citet{Howarth1989}.  Note that the measured $V \sin i$ 
is very large for all the targets except HD~155806, but given the presence of disk-like 
emission in its spectrum, we expect that it is a rapid rotator viewed from a lower 
inclination angle with smaller projected rotational Doppler shifts  \citep{Negueruela2004}.
 
\placetable{tab2}      
\begin{deluxetable*}{lclccccccc}
\tabletypesize{\scriptsize}
\tablenum{2}
\tablecaption{Rapidly Rotating O-type Stars\label{tab2}}
\tablewidth{0pt}
\tablehead{
\colhead{Star} & 
\colhead{No.} & 
\colhead{Spectral} & 
\colhead{Ref.} &
\colhead{$V_r$} & 
\colhead{Ref.} &
\colhead{$V\sin i$} & 
\colhead{Ref.} & 
\colhead{$V_\infty$} & 
\colhead{$-\log (-dM/dt)$}   \\ 
\colhead{Name} & 
\colhead{IUE} & 
\colhead{Classification} & 
\colhead{Code} &
\colhead{(km s$^{-1}$)} & 
\colhead{Code} &
\colhead{(km s$^{-1}$)} & 
\colhead{Code} & 
\colhead{(km s$^{-1}$)} & 
\colhead{($M_\odot ~{\rm yr}^{-1}$)}  
}
\startdata
HD 46485             & \phn\phn  2 & O7 V ((f))n var? & 1 & \phs\phn    26  & \phn 4 &     342 & 12 &    2250 & 6.6 \\
HD 60848             & \phn     12 & O8: V:pe         & 1 & \phs\phn\phn 6  & \phn 5 &     231 & 12 &    1650 & 6.8 \\
HD 93521             &         150 & O9.5 IIInn       & 1 & \phn      $-14$ & \phn 6 &     341 & 12 &    1075 & 7.2 \\
HD 149757            &         108 & O9.2 IVnn        & 1 & \phs\phn    15  & \phn 7 &     348 & 12 &    1640 & 7.2 \\
HD 155806            & \phn\phn  4 & O7.5 V((f))(e)   & 1 & \phs\phn    10  & \phn 4 & \phn 92 & 12 &    2900 & 6.6 \\
HD 191423            & \phn\phn  1 & ON9 II-IIInn     & 1 & \phn      $-52$ & \phn 8 &     392 & 12 &    1350 & 6.6 \\
BD+34$^{\circ}$ 1058 & \phn\phn  1 & O8 Vn            & 2 & \phn      $-38$ & \phn 9 &     401 & 12 &    2315 & 6.3 \\
\tableline
VFTS 102             & \phn\phn  0 & O9: Vnnne+       & 3 & \phs       228  &     10 &     610 & 13 & \nodata & \nodata \\
VFTS 285             & \phn\phn  0 & O7.5 Vnnn        & 3 & \phs       228  &     11 &     609 & 13 &    1287 & \nodata \\
\enddata
\tablecomments{
References: 
 1. \citet{MaizApellaniz2013};
 2. \citet{Tovmassian1994};
 3. \citet{Walborn2014};
 4. \citet{Grunhut2017};
 5. \citet{Boyajian2007};
 6. \citet{Howarth1993};
 7. \citet{Reid1993};
 8. \citet{Mahy2013};
 9. This paper;
10. \citet{Dufton2011};
11. \citet{Sana2013};
12. \citet{Penny1996};
13. \citet{Ramirez2013}.
}
\end{deluxetable*}

We collected all the available spectra of the Galactic rapid rotators from 
the Mikulski Archive for Space Telescopes (MAST\footnote{http://archive.stsci.edu/iue/}) 
and transformed these into normalized flux on a uniform $\log \lambda$ grid.  If more than 
one spectrum was available, then we formed an average spectrum to improve the S/N ratio. 
Finally all the spectra (IUE and HST/COS) were smoothed to a resolving power of 
$R=7500$ for ease of comparison.  The spectra of the Galactic and LMC stars are collected 
in Figure~3 for both the \ion{N}{5} and \ion{Si}{4} transitions.  Each spectrum is plotted 
as a function of Doppler-shifted velocity in the target star's frame of reference  
according to the stellar radial velocity $V_r$ listed in Table~2. The stars are ordered from 
weak and fast winds at the top to strong and slow winds at the bottom.
We see that the \ion{N}{5} doublet absorption features are found to some extent 
in all these stars with the sole exception of VFTS~102.  The outflow velocities associated 
with these features are generally fast as expected for main sequence stars with large 
escape velocities.  However, the spectra of the two stars at the bottom of Figure~3, 
HD~191423 and HD~93521, both show slow apparent velocities in the deepest parts of the 
profile with decreasing absorption towards more negative velocities.  Some of these trends
are reflected in the \ion{Si}{4} lines.  Again, VFTS~102 shows no obvious wind features, 
while the lines appear as a composite of slow and fast components in most of the other 
stars.  The strong absorption component found in the spectrum of VFTS~285 is also apparent
among the Galactic rapid rotators plotted in the lower part of Figure~3, and these strong
components are associated with slower outflow (as discussed for the case of HD~93521 by 
\citealt{Howarth1993}, \citealt{Bjorkman1994}, and \citealt{Massa1995}).  
In the next subsection, we discuss how these properties may reflect changes in the 
wind properties with stellar co-latitude.  

\begin{figure*}
\begin{center} 
{\includegraphics[height=20cm,trim= 0.7in 0.0in 0.0in 0.0in]{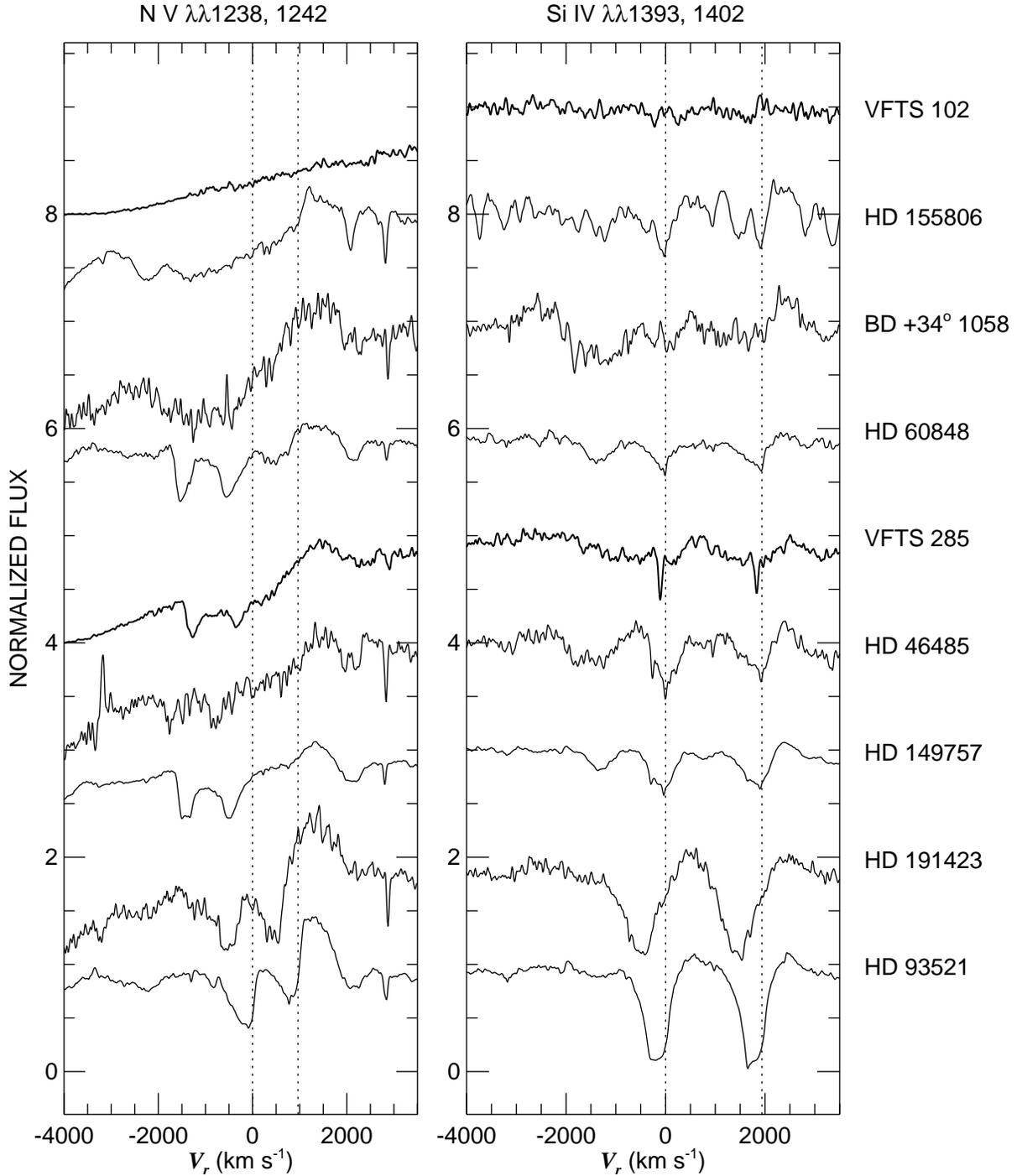}} 
\end{center} 
\caption{The \ion{N}{5} $\lambda\lambda 1238, 1242$ and \ion{Si}{4} $\lambda\lambda 1393, 1402$
wind line profiles for a sample of IUE spectra of Galactic, rapidly rotating, O-stars 
compared to HST/COS spectra of LMC stars VFTS~102 and VFTS~285.  Each spectrum is 
plotted as a function of radial velocity of the blue component of the doublet in the 
reference frame of the star, and they are offset by integer units of normalized flux
for ease of comparison.  The rest velocities of each doublet component are indicated by
vertical dotted lines.  
} 
\label{fig3}
\end{figure*}

\placefigure{fig3}     

\subsection{VFTS 285}                             

The \ion{N}{5} $\lambda\lambda 1238, 1242$ doublet appears in the red 
wing of Ly$\alpha$ (Fig.~3), and it clearly displays the blue-shifted absorption and 
red-shifted emission characteristic of wind lines.  We measured the radial velocity
of the two absorption minima as $V_r = -1059 \pm 20$ km~s$^{-1}$. Adopting a  
stellar velocity of $V_r = +228 \pm 7$ km~s$^{-1}$ \citep{Sana2013} leads to a wind terminal 
velocity of $v_\infty = 1287 \pm 21$ km~s$^{-1}$ in the star's frame, which is comparable 
to the average of $v_\infty = 1320 \pm 315$ km~s$^{-1}$ for O7-8~V stars in the 
30~Dor region of the LMC (see Table~6 in \citealt{Crowther2016}).  On the other
hand, the \ion{Si}{4} $\lambda\lambda 1393, 1402$ profile shows shallow 
and weak absorption that extends to similar blue-shift, even weaker red-shifted
emission, and a deep and narrow absorption at a smaller outflow velocity.  
The minima of the latter component yield a velocity of $V_r = +121 \pm 10$ km~s$^{-1}$, 
implying a slower wind outflow velocity of $v_\infty = 107 \pm 12$ km~s$^{-1}$ 
in the frame of reference of the star. 
The deep absorption component has a minimum flux that is only $\approx 12\% $ of the continuum
and a full-width at half-maximum (FWHM) of $\approx 60$ km~s$^{-1}$ in the original COS spectrum 
before the ISM-component removal and smoothing.  This FWHM is about $20\times$ smaller than 
that associated with rotational broadening ($2~V\sin i$), so the feature does not form in 
the stellar photosphere.  Instead, we suggest that presence of both broad and narrow 
features in \ion{Si}{4} is indicative of two components in the stellar wind. 

We suggest that the differences between the \ion{N}{5} and \ion{Si}{4} profiles 
result from a variation in the wind outflow between the poles and equator
and subsequent ionization changes. 
\citet{Gagnier2019a} present models for the wind variations with colatitude 
for very rapidly rotating stars that are based upon physically realistic models
of the stellar structure and surface temperature variation.   We show an 
example of the shape of a rapidly rotating star in Figure~4 for the case 
where the ratio of equatorial velocity to Keplerian velocity at the equator 
is $\omega = 0.992$.  The shape is defined by the Roche model in which the 
stellar mass is assumed to be a point source at the center and the outer 
boundary is set to an equipotential surface defined by the gravitational 
and centrifugal forces.  The temperature varies from hot at the pole 
to cool at the equator ($T_{eq}= 0.55 T_{pole}$ for $\omega = 0.992$ 
according to equation 32 of \citealt{EspinosaLara2011}). The mass loss rate 
reaches a local maximum at the poles due to the intense flux at that point, 
and declines towards the equator. However, 
\citet{Gagnier2019a} show that if the gas temperature in the region where the 
wind originates drops below the bi-stability jump at $T \approx 25000$~K, then 
lower ionization transitions emerge that can act to create a sharp rise in 
the radiation-driven wind.  The result is a two-wind regime in which 
significant mass loss occurs both at the poles (fast and low density) and 
at the equator (slow and dense). In the fastest rotators that span 
the bi-stability jump, the equatorial outflow dominates. 
 
\begin{figure*}
\begin{center} 
{\includegraphics[width=15cm,angle=0,trim= 0.0in 1.0in 0.2in 1.0in]{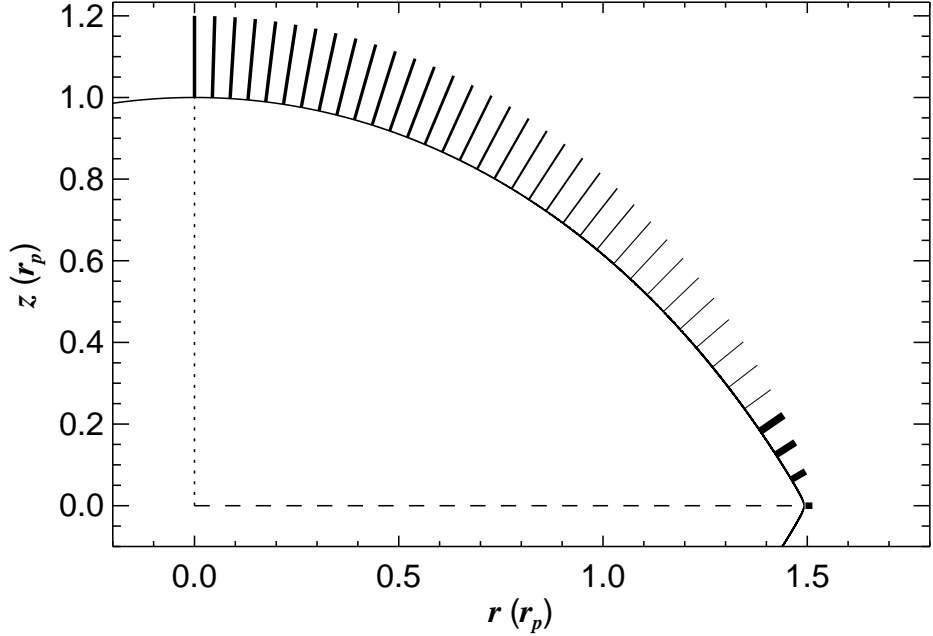}} 
\end{center} 
\caption{The solid line represents a quarter cross section of 
a very rapidly rotating star like VFTS~285 for an angular rotation  
rate of $\omega=0.992$ relative to the critical rate.  
The figure is plotted in terms of cylindrical coordinates for the 
radial $r$ and polar $z$ directions in units of the polar radius $r_p$.  
The dashed and dotted lines represent the equatorial plane and polar axis, respectively. 
The vectors placed at various surface normals represent the wind velocity
(length) and local mass loss rate (thickness) for the two-wind model 
from \citet{Gagnier2019a}.  We surmise that the deep, slow absorption 
in the \ion{Si}{4} profile originates in the strong outflow around the 
equator and that the faster outflow observed in both the \ion{N}{5} and 
\ion{Si}{4} profiles forms in the regions closer to the pole.} 
\label{fig4}
\end{figure*}

\placefigure{fig4}     

We suggest that the deep and narrow component of the \ion{Si}{4} profile 
originates in the slow and dense outflow from the equator and that the
faster, blue-shifted absorption observed in the \ion{N}{5} profile 
forms in the polar winds, which are more comparable to those of 
slowly rotating main sequence O-stars.   We suspect that the slower, 
deep absorption component is absent from the \ion{N}{5} profile because the 
equatorial region gas is too dense and cool to create highly ionized N. 
Conversely, the faster, polar outflow is only weakly present in 
the \ion{Si}{4} profile because Si is ionized beyond Si$^{+3}$
in the rarefied and hotter gas of the polar regions.  

It is instructive to compare the \ion{Si}{4} profile of VFTS~285 to those 
model profiles presented by \citet{Bjorkman1994} for the case of the
rapid rotator HD~93521.  In Section 4.2 of their paper, \citet{Bjorkman1994} 
show how changes in a given single model parameter from the default set 
lead to shape differences in the wind line.  The transition from a low
velocity, deep absorption (equatorial outflow) to a high velocity, 
weaker absorption (polar outflow) creates a mid-range, flat ``shelf''
in the profile that appears in most of the model profiles.  
The model profile that most closely resembles the observed \ion{Si}{4} profile
is that in their Figure~11 for a total polar optical depth of $T=1.0$, 
which is less than their default value of $T=3.3$ for the \ion{Si}{4} profile
formed in the wind of HD~93521.  This difference is probably due to 
the hotter photospheric temperature of VFTS~285 that will tend to 
over-ionize Si and reduce the optical depth of \ion{Si}{4} in the polar regions.
The other difference from the \citet{Bjorkman1994} models is the observed very 
narrow appearance of the low velocity absorption that we expect is related
to the very low effective gravity at the equator and subsequent small 
terminal velocity (see below).  An alternative explanation for the appearance
of the \ion{Si}{4} profile might be that we are simply observing a more or less
spherical wind in which the optical depth is too low at large distance 
from the star to create significant absorption at greater blue-shift. 
However, in this case we would expect a smooth transition in profile shape 
from the low velocity deeper core to the high velocity extension 
(see, for example, the low optical depth model profiles presented by 
\citealt{Castor1979} in their Fig.~5).  Instead, the sudden transition 
from steep to shallow slope evident in the profile of \ion{Si}{4} is 
inconsistent with such spherical models, but is the predicted outcome 
for two-component winds. 

\citet{Dwarkadas2002} show how the terminal wind velocity 
should vary with the local escape velocity at colatitude $\theta$ according 
to (their equation 8)
$${{v_\infty(\theta)}\over{v_\infty(0)}} = 
\Bigg[{{g_{\rm eff}(\theta)}\over{g_{\rm eff}(0)}}\Bigg]^{1/2}$$
where $g_{\rm eff}$ is the local effective gravity.  If we suppose that 
the measured outflow velocities for \ion{N}{5} and \ion{Si}{4}
correspond to $v_\infty(0)$ and $v_\infty(90)$, for the pole
and equator respectively, then the relation above can be solved to find 
$${{v_\infty(90)}\over{v_\infty(0)}} = 
\Bigg( {r_p \over r_e} \Bigg) (1 - \omega^2)^{1/2}$$
where $r_p$ and $r_e$ are the polar and equatorial radii and 
$$\omega^2 = {{\Omega^2} \over {GM/r_e^3}}.$$
Then for near-critical rotation in which $r_p/r_e = 2/3$ \citep{EspinosaLara2011}, 
we find a normalized rotation frequency of $\omega = 0.992 \pm 0.002$, i.e., 
VFTS~285 is rotating very close to the critical rate of $\omega = 1$.
Note that this result relies on the assumption that the observed wind velocity 
measures $v_\infty$. If, for example, line formation occurs predominantly in 
denser gas closer to the star where the flow has not reached terminal velocity,
then our estimate of $\omega$ would be an upper limit.   

The extremely large projected rotational velocity of VFTS~285 also 
suggests near critical rotation.  \citet{SabinSanjulian2017} estimate
that the star has a mass $M/M_\odot=20.1 \pm 0.9$ and radius $R/R_\odot=6.60 \pm 0.75$. 
The projected rotational velocity is so large that it is safe to 
assume that the inclination is $i\approx 90^\circ$ and $\sin i \approx 1$, 
and thus $V\sin i \approx \Omega r_e = 1.5 \Omega r_p$.  Setting the 
adopted radius to $r_p$ then yields $\Omega = (8.8 \pm 1.1) \times 10^{-5}$ s$^{-1}$
and a rotation period $P = 2 \pi / \Omega = 19.7 \pm 2.5$~h.  The Keplerian angular 
rotational velocity is $\Omega_K = \sqrt{GM/r_e^3} = (9.0 \pm 1.6) \times 10^{-5}$ s$^{-1}$, 
so the estimated ratio is $\omega = \Omega / \Omega_K = 0.98 \pm 0.08$. 
This is the same within uncertainties as that from the previous paragraph found by 
comparing the difference in wind speed between pole and equator.  Thus, VFTS~285 must 
represent a case of a star very close to critical rotation and oriented with 
its equator directly along the line of sight ($i\approx 90^\circ$), 
which allows the slow, dense equatorial wind, viewed projected against the star, 
to appear in the low velocity, blue-shifted absorption component of the \ion{Si}{4} profile.  
 
\subsection{VFTS 102}                             

There is no evidence of a P~Cygni wind profile for either the 
\ion{N}{5} or \ion{Si}{4} transitions in the spectrum of VFTS~102 (Fig.~3). 
This is not unexpected given the O9~V classification of the star. 
The cooler, late subtype O-stars show little evidence of wind 
features in these transitions among main sequence stars in the Galaxy
\citep{Walborn1985, Howarth1989} and in the LMC \citep{Crowther2016}, 
due to their relatively lower mass loss rates and cooler photospheres. 
However, the O9~V stars usually show photospheric components of \ion{Si}{4}
which are very weak or absent in the spectrum of VFTS~102.  We suspect 
that their absence is partially due to the extremely large projected rotational 
velocity that causes all the photospheric lines to appear very broad 
and shallow.  It is also possible that our ISM line removal scheme may 
have inadvertently subtracted out any minor \ion{Si}{4} components from 
the star.  The lack of deep and narrow \ion{Si}{4} components like those 
observed in the spectrum of VFTS~285 is puzzling given the rapid rotation 
of VFTS~102. However, since VFTS~102 is cooler, its equatorial outflow may 
only sustain lower ionization stages of Si, rendering it invisible in 
the \ion{Si}{4} transitions. 
A similar situation exists in the cooler Be stars with outflowing disks. 
\citet{Grady1987} found that wind components of \ion{Si}{4} are sometimes
observed in rapidly rotating Be stars (see below), but these are rare in those 
Be stars with high inclination presumably because their dense, cooler disks, 
seen in projection against the star, harbor lower ionization states. 

Evidence that the star is losing mass from the equatorial region comes
from the Balmer emission lines that place VFTS~102 in the Oe category. 
We show in Figure~5 the H$\alpha$ emission line as observed with the 
VLT X-shooter spectrograph \citep{Vernet2011} on 2013 November 9. 
The line appears double-peaked with a central narrow emission component 
that results from incomplete removal of the surrounding nebular emission. 
Such double peaks are generally interpreted as forming in a circumstellar 
decretion disk like those found around the classical Be stars \citep{Rivinius2013}.
The H$\alpha$ emission line is particularly strong in the spectrum 
of VFTS~102 with an equivalent width of $-42 \pm 2$ \AA\ and emission peaks 
reaching to four times the continuum flux level above the continuum
(after removal of the nebular component).  We used the equivalent width and 
the method presented by \citet{Grundstrom2006} to derive a disk half-light radius for 
H$\alpha$ emission that is approximately nine times larger than the stellar mean radius, 
although we caution that this estimate is based upon an extrapolation to 
higher temperature of model results for Galactic abundance stars.  
In any case, the extremely large H$\alpha$ emission strength demonstrates 
that the star is losing mass preferentially through an equatorial flow into 
a dense and extended circumstellar disk. 
 
\begin{figure*}
\begin{center} 
{\includegraphics[height=7cm,angle=0]{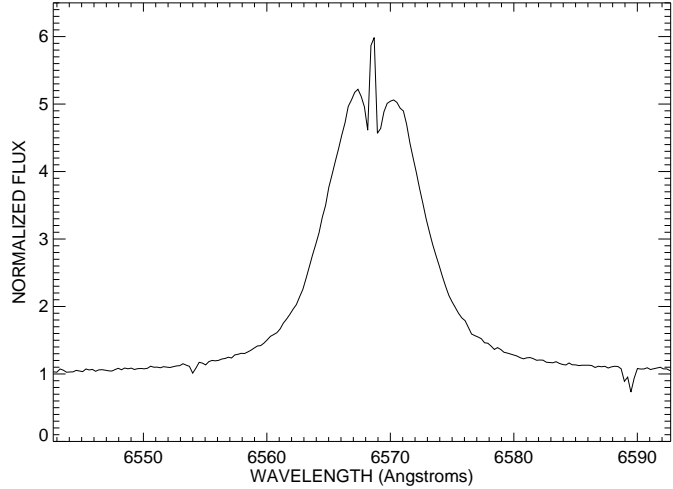}} 
\end{center} 
\caption{X-shooter spectrum of the H$\alpha$ profile of VFTS~102 that 
shows double-peaked emission from a large circumstellar disk.} 
\label{fig5}
\end{figure*}

\placefigure{fig5}     

 
\section{Conclusions}                               
 
The LMC stars VFTS~102 and 285 are the record holders for fastest equatorial 
rotational velocity, and the HST/COS spectra of these stars offers us
a rare opportunity to observe how rapid rotation affects their radiatively 
driven mass loss.  The wind lines in the spectrum of VFTS~285 show evidence
of both a fast outflow that is typical of main sequence O-stars and a striking 
and unusual slow component visible in the \ion{Si}{4} $\lambda\lambda 1393, 1402$
profiles.  We argue that these features support the two-wind regime model
for very rapid rotators presented by \citet{Gagnier2019a}.  In this model 
the wind mass loss rate is high and the speed fastest in the polar regions 
where the stellar temperature and local escape velocity reaches maximum.
However, in the case of very rapid rotation, the equatorial zone becomes 
cool enough that the atmospheric temperature falls below the bi-stability jump
where new, low ionization absorbers help drive a dense and slow outflow. 
We show that there are similar slow outflow components in the \ion{Si}{4}
wind lines of some Galactic rapid rotators, such as HD~93521 and HD~191423, 
although their outflows are not as slow as that in VFTS~285, presumably as a 
result of their slower rotation rates and the smaller difference between the 
gravity at the poles and equator.

There is little or no evidence of P~Cygni type wind profiles for 
the cooler star VFTS~102, which is likely due to the lower ionization 
conditions of its photosphere.  On the other hand, there is strong 
and double-peaked H$\alpha$ emission in the optical spectrum that 
demonstrates the presence of an extended circumstellar disk that 
is probably the result of mass loss from the equatorial region which 
carries away the highest amount of angular momentum.   The fact that a large disk 
is present in the cooler star VFTS~102 but absent from the hotter VFTS~285
supports the idea that the intense flux of hotter stars tends to 
ablate away gas in a circumstellar disk \citep{Kee2016}.

These two cases show that although a two-wind regime may exist in a 
very rapidly rotating star, observing the spectral signature of the slower
equatorial outflow depends upon several conditions: 
the polar temperature, 
how close the star is to critical rotation ($\omega = 1$), 
the spatial variation of ionization states in the wind,
and the inclination of the spin axis to our line of sight. 
The diversity of wind profiles found among Galactic rapid rotators (Fig.~3) 
reflects the differences among these parameters in different stars.   
We encourage the development of three dimensional radiative transfer 
codes (e.g., \citealt{Hennicker2018}) in order to compute detailed line profiles for 
stars with wind outflows and ionization levels dependent on co-latitude and 
to test the predictions of the two-wind regime model.  It is important to determine 
the role of equatorial mass loss, because such an outflow removes relatively more
angular momentum and can alter the rotational evolution of massive stars 
\citep{Gagnier2019b}.

 
\acknowledgments 
 
We are grateful to Nolan Walborn (deceased) and Denise Taylor of STScI 
for their aid in planning the observations with HST.
Support for Program number GO-14246 was provided by NASA through a grant from the 
Space Telescope Science Institute, which is operated by the Association of Universities 
for Research in Astronomy, Incorporated, under NASA contract NAS5-26555. 
Some of the data presented in this paper were obtained from the 
Mikulski Archive for Space Telescopes (MAST).  Support for MAST for 
non-HST data is provided by the NASA Office of Space Science via grant 
NNX13AC07G and by other grants and contracts.
Institutional support was provided from the GSU College of Arts and Sciences. 


\facilities{HST (COS), VLT:Kueyen (X-shooter)}


\bibliographystyle{aasjournal}
\bibliography{paper}{}

\end{document}